\documentclass[12pt]{article}
\usepackage{epsfig}
\renewcommand{\bar}[1]{\overline{#1}}
\newcommand{\qu}{{\rm q}}
\newcommand{\qb}{${\rm\bar q}$}
\newcommand{\pvec}{\vec p}
\newcommand{\kvec}{\vec k}
\newcommand{\rvec}{\vec r}
\newcommand{\Rvec}{\vec R}
\newcommand{\ieps}{i\varepsilon}
\newcommand{\pl}{{||}}
\newcommand{\order}[1]{${ O}\left(#1 \right)$}
\newcommand{\eq}[1]{(\ref{#1})}
\newcommand{\half}{{$\frac{1}{2}$}} 
\newcommand{\cd}{\makebox[0.08cm]{$\cdot$}}

\begin {document}
\begin{flushright}
{\small
SLAC--PUB--10777\\
October 2004\\}
\end{flushright}

\begin{center}
{{\bf\LARGE Novel QCD Aspects of Hard Diffraction,\\
Antishadowing, and Single-Spin Asymmetries }\footnote{Work supported
by Department of Energy contract DE--AC02--76SF00515.}}

\bigskip
{\it Stanley J. Brodsky \\
Stanford Linear Accelerator Center \\
Stanford University, Stanford, California 94309 \\
E-mail:  sjbth@slac.stanford.edu}
\medskip
\end{center}

\vfill

\begin{center}
{\bf\large Abstract }
\end{center}

It is usually assumed -- following the parton model -- that the
leading-twist structure functions measured in deep  inelastic
lepton-proton scattering are simply the probability distributions
for finding quarks and gluons in the target nucleon.  In fact, gluon
exchange between the outgoing quarks and the target spectators
effects the leading-twist structure functions in a profound way,
leading to  diffractive leptoproduction processes, shadowing and
antishadowing of nuclear structure  functions, and target spin
asymmetries, physics not incorporated in the light-front
wavefunctions of the target computed in isolation. In particular,
final-state  interactions from gluon exchange lead to single-spin
asymmetries in  semi-inclusive deep inelastic lepton-proton
scattering which are not  power-law suppressed in the Bjorken limit.
The shadowing and antishadowing of nuclear structure functions in
the Gribov-Glauber picture is  due respectively to the destructive
and constructive interference of amplitudes arising from the
multiple-scattering of quarks in the nucleus.   The effective
quark-nucleon scattering amplitude includes Pomeron and Odderon
contributions from multi-gluon exchange as well as Reggeon
quark-exchange contributions.  Part of the anomalous NuTeV result
for $\sin^2\theta_W$ could be due to the non-universality  of
nuclear antishadowing for charged and neutral currents.  Detailed
measurements of the nuclear dependence of individual quark structure
functions are thus needed to establish the distinctive phenomenology
of shadowing and antishadowing and to make the NuTeV results
definitive.  I also discuss diffraction dissociation as a tool for
resolving hadron substructure Fock state by Fock state and for
producing leading heavy quark systems.

\vfill

\begin{center}
{\it Presented at the  \\
34th International Symposium on Multiparticle Dynamics (ISMD 2004)
\\
 Sonoma State University, Rohnert Park, California\\
July 26 - August 1, 2004. }\\
\end{center}

\vfill \newpage

\section{Light-Front Wavefunctions and Structure Functions}

The concept of a wave function of a hadron as a composite of relativistic quarks and
gluons is naturally formulated in terms of the light-front Fock expansion at fixed
light-front time, $\tau=x \cd \omega$~{Brodsky:1997de}.  The four-vector $\omega$, with
$\omega^2 = 0$, determines the orientation of the light-front plane; the freedom to
choose $\omega$ provides an explicitly covariant formulation of light-front
quantization~\cite{cdkm}. The light-front wave functions (LFWFs)
$\psi_n(x_i,k_{\perp_i},\lambda_i)$, with $x_i={k_i \cd \omega\over P\cd \omega}$,
$\sum^n_{i=1} x_i=1, $ $\sum^n_{i=1}k_{\perp_i}=0_\perp$, are the coefficient functions
for $n$ partons in the Fock expansion, providing a general frame-independent
representation of the hadron state.  Matrix elements of local operators such as spacelike
proton form factors can be computed simply from the overlap integrals of light front wave
functions in analogy to nonrelativistic Schr\"odinger theory. In principle, one can solve
for the LFWFs directly from the fundamental theory using methods such as discretized
light-front quantization, the transverse lattice, lattice gauge theory moments, or
Bethe--Salpeter techniques.  The determination of the hadron LFWFs from phenomenological
constraints and from QCD itself is a central goal of hadron and nuclear physics.

Ever since the earliest days of the parton model, it has been assumed that the
leading-twist structure functions $F_i(x,Q^2)$ measured in deep inelastic lepton
scattering are determined by the {\it probability} distributions of quarks and gluons as
determined by the light-front wave functions of the target.  For example, the quark
distribution is \begin{equation} { P}_{\qu/N}(x_B,Q^2)= \sum_n \int^{k_{iT}^2<Q^2}\left[
\prod_i\, dx_i\, d^2k_{T i}\right] |\psi_n(x_i,k_{T i})|^2 \sum_{j=q} \delta(x_B-x_j).
\end{equation} The identification of structure functions with the square of
light-front wave functions is usually made in the LC gauge, $\omega\cdot A = A^+=0$,
where the path-ordered exponential in the operator product for the forward virtual
Compton amplitude apparently reduces to unity.  Thus the deep inelastic lepton scattering
cross section appears to be fully determined by the probability distribution of partons
in the target.

\subsection{The Paradox of Diffractive Deep Inelastic Scattering}

A remarkable feature of deep inelastic lepton-proton scattering at HERA is that
approximately 10\% events are
diffractive~\cite{Abramowicz:1999eq,Adloff:1997sc,Breitweg:1998gc}: the
target proton remains intact and there is a large rapidity gap between the
proton and the
other hadrons in the final state.  These diffractive deep inelastic
scattering (DDIS)
events can be understood most simply from the perspective of the color-dipole
model~\cite{Raufeisen:2000sy}: the $q \bar q$ Fock state of the high-energy
virtual photon
diffractively dissociates into a diffractive dijet system.  The
color-singlet exchange of
multiple gluons  between  the color dipole of the $q \bar q$ and the quarks
of the target
proton leads to the diffractive final state.  The same hard pomeron
exchange also controls
diffractive vector meson electroproduction at large photon
virtuality~\cite{Brodsky:1994kf}.
One can show by analyticity and crossing symmetry that amplitudes with
$C=+$ hard-pomeron exchange have a nearly imaginary phase.

This observation presents a paradox:
deep inelastic scattering is usually discussed in terms of the
parton model.  If one chooses the conventional parton model frame where the
photon
light-front momentum is negative
$q+ = q^0 + q^z  < 0$, then the virtual photon cannot produce a virtual $q
\bar q$ pair.
Instead, the virtual photon always interacts with a quark constituent with
light-cone
momentum fraction
$x = {k^+\over p^+} = x_{bj}.$  If one chooses light-cone gauge $A^+=0$,
then the gauge
link associated with the struck quark (the  Wilson line)  becomes unity.
Thus the
struck ``current" quark experiences no final-state interactions.  The
light-front
wavefunctions $\psi_n(x_i,k_{\perp i}$ of the proton which determine the
quark probability
distributions
$q(x, Q)$ are real since the proton is stable.  Thus it appears impossible
to generate the
required imaginary phase, let alone the large rapidity gaps associated with  of
DDIS.

This paradox was resolved by Paul Hoyer, Nils Marchal, Stephane Peigne, Francesco Sannino
and myself~\cite{Brodsky:2002ue}.  It is helpful to consider the case where the virtual
photon interacts with a strange quark --  the $s \bar s$ pair is assumed to be produced
in the target by gluon splitting.  In the case of Feynman gauge, the struck $s$ quark
continues to interact in the final state via gluon exchange as described by the Wilson
line.  The final-state interactions occur at a light-cone time $\Delta\tau \simeq 1/\nu$
after the virtual photon interacts with the struck quark.   When one integrates over the
nearly-on-shell intermediate state, the amplitude acquires an imaginary part. Thus the
rescattering of the quark produces a separated color-singlet $s \bar s$ and an imaginary
phase.

In contrast, in the case of the light-cone gauge $A^+ = \omega \cdot A =0$, one must
consider the final state interactions of the (unstruck) $\bar s$ quark.  Light-cone gauge
is singular---in particular, the gluon propagator \begin{equation} d_{LC}^{\mu\nu}(k) =
\frac{i}{k^2+\ieps}\left[-g^{\mu\nu}+\frac{\omega^\mu k^\nu+ k^\mu
\omega^\nu}{\omega\cdot k}\right] \label{lcprop} \end{equation} has a pole at $k^+ = 0$
which requires an analytic prescription.  In final-state scattering involving nearly
on-shell intermediate states, the exchanged momentum $k^+$ is of \order{1/\nu} in the
target rest frame, which enhances the second term in the propagator.  This enhancement
allows rescattering to contribute at leading twist even in LC gauge.  Thus the
rescattering contribution survives in the Bjorken limit because of the singular behavior
of the propagator of the exchanged gluon at small $k^+$ in $A^+=0$ gauge.    The net
result is  gauge invariant and identical to the color dipole model calculation.

\begin{figure}[htb]
\begin{center}
\leavevmode {\epsfysize=4cm \epsffile{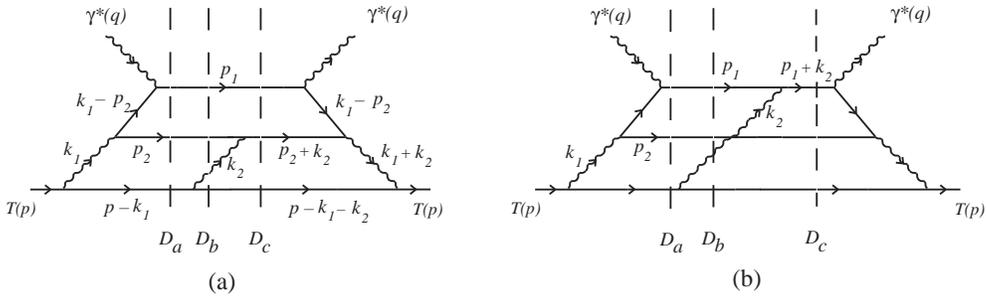}}
\end{center}
\caption[*]{\baselineskip 13pt Two types of final state interactions.  (a)
Scattering of
the antiquark ($p_2$ line), which in the aligned jet kinematics is part of
the target
dynamics.  (b) Scattering of the current quark ($p_1$ line).  For each
light-front
time-ordered diagram, the potentially on-shell intermediate
states---corresponding to the
zeroes of the denominators $D_a, D_b, D_c$---are denoted by dashed lines.}
\label{brodsky1}
\end{figure}

The calculation of the rescattering effects on DIS in Feynman and
light-cone gauge through three loops is given in detail for a simple
Abelian model in
Ref.~\cite{Brodsky:2002ue}.  Figure~\ref{brodsky1} illustrates two LCPTH
diagrams which contribute to the forward $\gamma^* T \to \gamma^*
T$ amplitude, where the target $T$ is taken to be a single quark.
In the aligned jet kinematics the virtual photon fluctuates into a
\qu\qb\ pair with limited transverse momentum, and the (struck)
quark takes nearly all the longitudinal momentum of the photon.
The initial \qu\ and \qb\ momenta are denoted $p_1$ and $p_2-k_1$,
respectively.  The result is
most easily expressed in eikonal form in terms of transverse
distances $r_T, R_T$ conjugate to $p_{2T}, k_T$.  The DIS cross
section can be expressed as
\begin{equation}
Q^4\frac{d\sigma}{dQ^2\, dx_B} = \frac{\alpha_{\rm
em}}{16\pi^2}\frac{1-y}{y^2} \frac{1}{2M\nu} \int
\frac{dp_2^-}{p_2^-}\,d^2\rvec_T\, d^2\Rvec_T\, |\tilde M|^2
\label{transcross} \end{equation} where \begin{equation} |\tilde{
M}(p_2^-,\rvec_T, \Rvec_T)| = \left|\frac{\sin \left[g^2\,
W(\rvec_T, \Rvec_T)/2\right]}{g^2\, W(\rvec_T, \Rvec_T)/2}
\tilde{A}(p_2^-,\rvec_T, \Rvec_T)\right| \label{Interference}
\end{equation}
is the resummed result.  The Born amplitude is
\begin{equation} \tilde
A(p_2^-,\rvec_T, \Rvec_T) = 2eg^2 M Q p_2^-\, V(m_\pl r_T)
W(\rvec_T, \Rvec_T) \label{Atildeexpr} \end{equation}
where $m_\pl^2 = p_2^-Mx_B + m^2 \label{mplus}$ and
\begin{equation}
V(m\, r_T) \equiv \int \frac{d^2\pvec_T}{(2\pi)^2}
\frac{e^{i\rvec_T\cdot\pvec_{T}}}{p_T^2+m^2} =
\frac{1}{2\pi}K_0(m\,r_T). \label{Vexpr} \end{equation}
The
rescattering effect of the dipole of the \qu\qb~ is controlled by
\begin{equation} W(\rvec_T, \Rvec_T) \equiv \int \frac{d^2\kvec_T}{(2\pi)^2}
\frac{1-e^{i\rvec_T\cdot\kvec_{T}}}{k_T^2}
e^{i\Rvec_T\cdot\kvec_{T}} = \frac{1}{2\pi}
\log\left(\frac{|\Rvec_T+\rvec_T|}{R_T} \right). \label{Wexpr}
\end{equation}
The fact that the coefficient of $\tilde A$ in
Eq.~\eq{Interference} is less than unity for all $\rvec_T,
\Rvec_T$ shows that the rescattering corrections reduce the cross
section in  analogy  to nuclear shadowing.

\vspace{0.3cm}

A new understanding of the role of final-state interactions in
deep inelastic scattering has thus
emerged.  The final-state interactions from
gluon exchange occurring immediately after the interaction of the
current produce a leading-twist diffractive component to
deep inelastic scattering $\ell p \to \ell^\prime p^\prime X$
due to the color-singlet exchange with the target system.  This rescattering
is described in the Feynman gauge by the path-ordered exponential (Wilson
line) in the
expression for the parton distribution function of the target.  The
multiple  scattering of the struck parton via instantaneous interactions
in the target generates dominantly imaginary diffractive amplitudes,
giving rise to an effective ``hard pomeron" exchange.  The presence of a
rapidity gap between the target and diffractive system requires that the
target remnant emerges in a color-singlet state; this is made
possible in any gauge by the soft rescattering of the final-state $s -\bar
s$ system.

Rikard Enberg, Paul Hoyer, Gunnar Ingelman and I have recently discussed
further
aspects of the QCD
dynamics of diffractive deep inelastic scattering~\cite{Brodsky:2004hi}.
We show that the quark structure function of the effective hard pomeron
has the same
form as the quark contribution of the gluon structure
function.
The hard pomeron  is not an
intrinsic part of the proton; rather it must be considered as  a dynamical
effect of the
lepton-proton interaction.

Our QCD-based picture also applies to diffraction in hadron-initiated
processes.  The
rescattering is different in virtual photon- and hadron-induced processes
due to the
different color environment, which accounts for the  observed
non-universality of
diffractive parton distributions.  In the hadronic case the color flow at
tree level can
involve color-octet as well as color-triplet separation.  Multiple
scattering of the
quarks and gluons can set up a variety of different color singlet domains.  This
framework also provides a theoretical basis for the phenomenologically
successful Soft
Color Interaction (SCI) model which includes rescattering effects and thus
generates a
variety of final states  with rapidity gaps.

As I review below, the final-state interactions from
gluon exchange between the outgoing quarks and the target spectator
system also lead to
single-spin asymmetries in semi-inclusive
deep inelastic lepton-proton scattering  which  are not
power-law suppressed at large photon virtuality $Q^2$ at fixed
$x_{bj}$~\cite{Brodsky:2002cx}

\subsection{The Origin of Nuclear Shadowing and Antishadowing}

The physics of nuclear shadowing in deep inelastic scattering
can be most easily understood in the laboratory frame using the
Glauber-Gribov picture~\cite{Glauber:1955qq,Gribov:1968gs,Gribov:1968jf}.  The
virtual photon, $W,$ or $Z^0$  produces a quark-antiquark
color-dipole pair which can interact diffractively or
inelastically on the nucleons in the nucleus.  The destructive
interference of diffractive amplitudes from pomeron exchange on
the upstream nucleons then causes shadowing of the virtual photon
interactions on the back-face
nucleons~\cite{Stodolsky:1966am,Brodsky:1969iz,Brodsky:1990qz,Ioffe:1969kf,
Frankfurt:1988zg,Kopeliovich:1998gv,Kharzeev:2002fm}.
The Bjorken-scaling diffractive interactions
on the nucleons in a nucleus
thus leads to the shadowing (depletion at small $x_{bj}$) of the nuclear
structure functions.

\begin{figure}[htb]
\begin{center}
\leavevmode {\epsfysize=4cm \epsffile{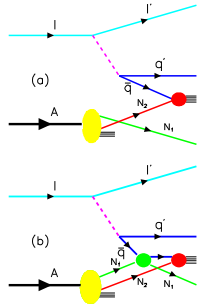}}
\end{center}
\caption[*]{\baselineskip 13pt The one-step and two-step processes in DIS on a nucleus.
If the scattering on nucleon $N_1$ is via pomeron exchange, the one-step and two-step
amplitudes are opposite in phase, thus diminishing the $\bar q$ flux reaching $N_2.$ This
causes shadowing of the charged and neutral current nuclear structure functions.
\label{bsy1f1}}
\end{figure}

As
emphasized by Ioffe~\cite{Ioffe:1969kf}, the coherence between
processes which occur on different nucleons at separation $L_A$
requires small Bjorken $x_{B}:$ $1/M x_B = {2\nu/ Q^2}  \ge L_A .$
The coherence between different quark processes is also the basis
of saturation phenomena in DIS and other hard QCD reactions at
small $x_B$~\cite{Mueller:2004se}, and coherent multiple parton scattering
has been used in the analysis of $p+A$ collisions in terms of the perturbative
QCD factorization approach~\cite{Qiu:2004da}.  An example of the interference
of one- and two-step processes in deep inelastic lepton-nucleus
scattering illustrated in Fig.~\ref{bsy1f1}.

\vspace{0.3cm}

An important aspect of the shadowing phenomenon is that the
diffractive contribution $\gamma^* N \to X N^\prime$ to deep
inelastic scattering (DDIS) where the nucleon $N_1$ in
Fig.~\ref{bsy1f1} remains intact is a constant fraction of the
total DIS rate, confirming that it is a leading-twist
contribution.  The Bjorken scaling of DDIS has been observed at
HERA~\cite{Adloff:1997sc,Martin:2004xw,Ruspa:2004jb}.  As shown in
Ref.~\cite{Brodsky:2002ue}, the leading-twist contribution to DDIS
arises in QCD in the usual parton model frame when one includes
the nearly instantaneous gluon exchange final-state interactions
of the struck quark with the target spectators.  The same final
state interactions also lead to leading-twist single-spin
asymmetries in semi-inclusive DIS~\cite{Brodsky:2002cx}.  Thus the
shadowing of nuclear structure functions is also a leading-twist
effect.

It was shown in Ref.~\cite{Brodsky:1989qz}  that if one allows for
Reggeon exchanges which leave a nucleon intact,  then one can
obtain {\it constructive} interference among the multi-scattering
amplitudes in the nucleus.   A Bjorken-scaling contribution to
DDIS from Reggeon exchange has in fact also been observed at
HERA~\cite{Adloff:1997sc,Ruspa:2004jb}.  The strength and energy
dependence of the $C=+$ Reggeon $t-$channel exchange contributions
to virtual Compton scattering is constrained by the
Kuti-Weisskopf~\cite{Kuti:1971ph} behavior $F_2(x) \sim
x^{1-\alpha_R}$ of the non-singlet electromagnetic structure
functions at small $x$.  The phase of the Reggeon exchange
amplitude is determined by its signature factor.  Because of this
phase structure~\cite{Brodsky:1989qz}, one obtains constructive
interference and {\it antishadowing} of the nuclear structure
functions in the range $0.1 < x < 0.2$ -- a pronounced excess of
the nuclear cross section with respect to nucleon
additivity~\cite{Arneodo:1992wf}.

In the case where the diffractive amplitude on $N_1$ is imaginary,
the two-step process has the phase $i \times i = -1 $ relative to
the one-step amplitude, producing destructive interference.  (The
second factor of $i$ arises from integration over the quasi-real
intermediate state.)  In the case where the diffractive amplitude
on $N_1$ is due to $C=+$ Reggeon exchange with intercept
$\alpha_R(0) = 1/2$, for example, the phase of the two-step
amplitude is ${1\over \sqrt 2}(1-i) \times i = {1\over \sqrt 2}
(i+1)$ relative to the one-step amplitude, thus producing
constructive interference and antishadowing.

The effective
quark-nucleon scattering amplitude includes Pomeron and Odderon
contributions from multi-gluon exchange as well as Reggeon quark-exchange
contributions~\cite{Brodsky:1989qz}.  The coherence of these
multiscattering nuclear processes leads to shadowing and antishadowing of
the electromagnetic nuclear structure functions in agreement with
measurements.  The Reggeon contributions to the quark scattering amplitudes
depend specifically on the quark flavor; for example the isovector
Regge trajectories couple differently to $u$ and $d$ quarks.  The
$s$ and $\bar s$ couple to yet different Reggeons.  This implies
distinct anti-shadowing effects for each quark and antiquark
component of the nuclear structure function.
Ivan Schmidt,  Jian-Jun Yang, and
I~\cite{Brodsky:2004bg} have shown that this picture leads to
substantially different antishadowing for charged and neutral current
reactions.

\begin{figure}[htb]
\begin{center}
\leavevmode {\epsfysize=10cm \epsffile{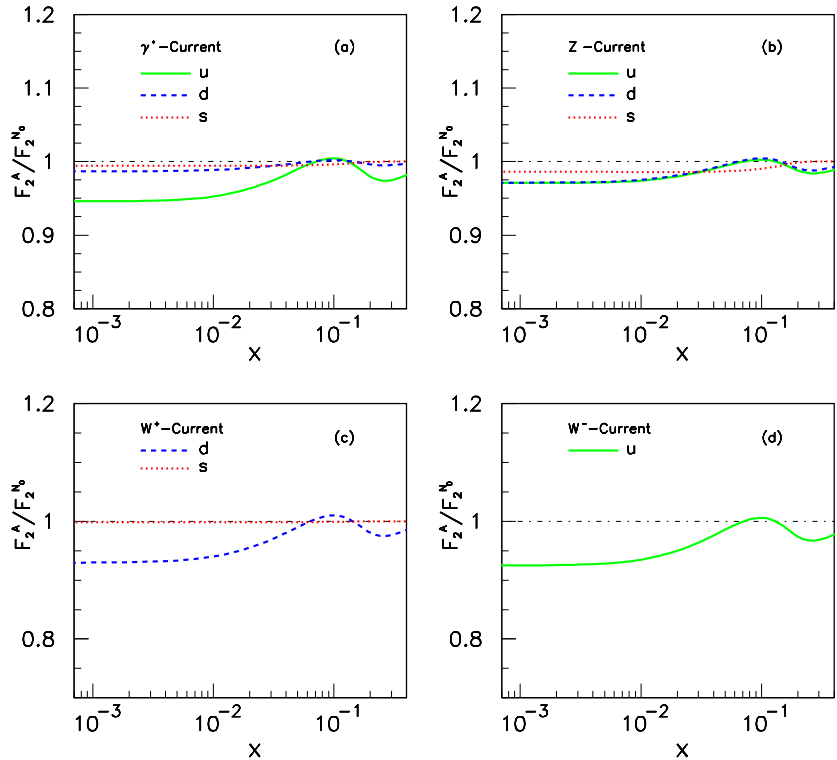}}
\end{center}
\caption[*]{\baselineskip 13pt
 The  quark contributions to the
ratios of structure functions at $ Q^2 = 1~\rm{GeV}^2$.  The solid,
dashed and dotted curves correspond to the $u$, $d$ and $s$ quark
contributions, respectively.  This corresponds in our model to the
nuclear dependence of the $\sigma(\bar u-A)$, $\sigma(\bar d-A)$,
$\sigma(\bar s-A)$ cross sections, respectively.  In order to
stress the individual contribution of quarks, the numerator of the
ratio $F_2^{A} / F_2^{N_0}$ shown in these two figures is obtained
from the denominator by a replacement $q^{N_0}$ into $q^{A}$ for
only the considered quark.  As a result, the effect of
antishadowing appears diminished.
 \label{bsy1f5}}
\end{figure}

\begin{figure}[ht]
\begin{center}
\leavevmode {\epsfysize=10cm \epsffile{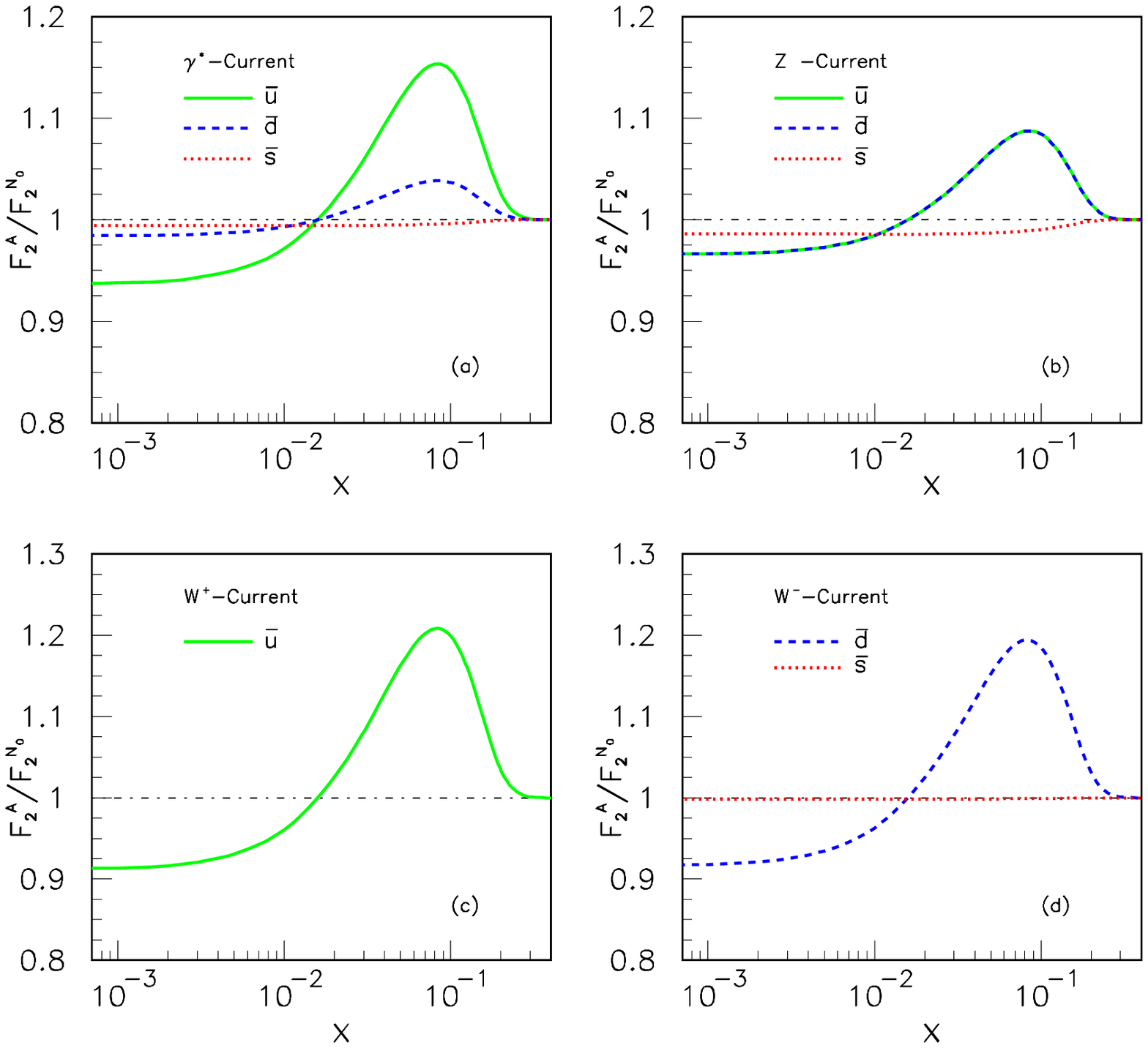}} 
\end{center}
\caption[*]{\baselineskip 13pt
 The anti-quark contributions to
ratios of the structure functions at $ Q^2 = 1~\rm{GeV}^2$.  The
solid, dashed and dotted curves correspond to $\bar{u}$, $\bar{d}$
and $\bar{s}$ quark contributions, respectively.  This corresponds
in our model to the nuclear dependence of the $\sigma(u-A)$,
$\sigma(d-A)$, $\sigma(s-A)$ cross sections, respectively.  In
order to stress the individual contribution of quarks, the
numerator of the ratio $F_2^{A} / F_2^{N_0}$ shown in these two
figures is obtained from the denominator by a replacement
$\bar{q}^{N_0}$ into $\bar{q}^{A}$ for only the considered
anti-quark.
 \label{bsy1f6}}
\end{figure}

Figs.~\ref{bsy1f5}--\ref{bsy1f6} illustrate the individual quark  $q$ and
anti-quark
$\bar{q}$ contributions to the ratio of the iron to nucleon structure functions
$R=F_2^{A} / F_2^{N_0}$ in a model calculation where the Reggeon
contributions are
constrained by the Kuti-Weisskopf behavior~\cite{Kuti:1971ph} of the
nucleon structure
functions at small $x_{bj}.$  Because the strange quark distribution is
much smaller than
$u$ and $d$ quark distributions, the strange quark contribution to the
ratio is very
close to 1 although $s^{A}/s^{N_0}$ may significantly deviate from 1.

\vspace{0.3cm}

\vspace{0.3cm}

Our analysis leads to substantially different nuclear antishadowing for
charged and
neutral current reactions; in fact, the neutrino and antineutrino DIS cross
sections are
each modified in different ways due to the various allowed Regge exchanges.  The
non-universality of nuclear effects will modify the extraction of the
weak-mixing angle
$\sin^2\theta_W$, particularly because of the strong nuclear effects for
the $F_3$
structure function.  The shadowing and antishadowing of the strange quark
structure
function in the nucleus can also be considerably different than that of the light quarks.
We thus find that part of the anomalous NuTeV result~\cite{McFarland:2003gx} for
$\sin^2\theta_W$ could be due to the non-universality  of nuclear
antishadowing for
charged and neutral currents.  Our picture also implies non-universality
for the nuclear
modifications of spin-dependent structure functions.

Thus the antishadowing of nuclear structure functions depends in detail on
quark flavor.
Careful measurements of the nuclear dependence of charged, neutral, and
electromagnetic
DIS processes are needed to establish the distinctive phenomenology of
shadowing and
antishadowing and to make the NuTeV results definitive.  It is also
important to map out
the shadowing and antishadowing of each quark component of the nuclear structure
functions to illuminate the underlying QCD mechanisms.  Such studies can be
carried out in
semi-inclusive deep inelastic scattering for the electromagnetic current at
Hermes and at
Jefferson Laboratory by tagging the flavor of the current quark or by using
pion and
kaon-induced Drell-Yan reactions.  A new determination of $\sin^2\theta_W$
is also
expected from the neutrino scattering experiment NOMAD at
CERN~\cite{Petti}.  A systematic
program of measurements of the nuclear effects in charged and neutral
current reactions
could also be carried out in high energy electron-nucleus colliders such as
HERA and
eRHIC, or by using high intensity neutrino beams~\cite{Geer}.

\subsection{Structure Functions Are Not Probability Functions}

As discussed above, the leading-twist contribution to DIS is affected by diffractive
rescattering of a quark in the target, a coherent effect which is not included in the
light-front wave functions computed in isolation, even in light-cone gauge.  Diffractive
contributions which leave the target intact do not resolve the quark structure of the
target, and thus there are contributions to structure functions which are not parton
probabilities~\cite{Brodsky:2002ue}.

The shadowing of nuclear structure functions is due to the destructive
interference
between rescattering amplitudes involving on-shell intermediate states with
a complex phase.
In contrast, the wave function of a stable target is strictly real since it
does not
have on-energy-shell intermediate state configurations.   The physics of
shadowing is thus not
included in the nuclear light-front wave functions, and a probabilistic
interpretation of the
DIS cross section is thus precluded.

As an alternative, one can augment the light-front wave functions with a
gauge link
corresponding to an external field created by the virtual photon $q \bar q$ pair
current~\cite{Belitsky:2002sm,Collins:2004nx}.  Such a gauge link is process
dependent~\cite{Collins:2002kn}, so the resulting augmented LFWFs are not
universal~\cite{Brodsky:2002ue,Belitsky:2002sm,Collins:2003fm}.  Such
rescattering
corrections are  not contained in the target light-front wave functions
computed in
isolation.

\section{ Single-Spin Asymmetries from Final-State
Interactions}

Spin correlations provide a remarkably sensitive window to
hadronic structure and basic mechanisms in QCD.   Among the most
interesting polarization effects are single-spin azimuthal
asymmetries  in semi-inclusive deep inelastic scattering,
representing the correlation of the spin of the proton target and
the virtual photon to hadron production plane: $\vec S_p \cdot
\vec q \times \vec p_H$~\cite{Avakian:2002td}.  Such asymmetries
are time-reversal odd, but they can arise in QCD through phase
differences in different spin amplitudes.

Until recently, the traditional explanation of pion electroproduction
single-spin asymmetries in semi-inclusive deep inelastic scattering is that
they are proportional to the transversity distribution of the quarks in
the hadron $h_{1}$~\cite{Jaffe:1996zw,Boer:2001zw,Boer:2002xc}
convoluted with the transverse momentum dependent fragmentation
(Collins) function $H^\perp_1$, the
distribution for a transversely polarized quark to fragment into
an unpolarized hadron with non-zero transverse momentum
\cite{Collins93,Barone:2001sp,Ma:2002ns,Goldstein:2002vv,Gamberg:2003ey}.

Dae Sung Hwang, Ivan Schmidt and I have showed that an alternative physical mechanism for
the azimuthal asymmetries  also exists~\cite{Brodsky:2002cx,Collins,Ji:2002aa}. The same
QCD final-state interactions (gluon exchange) between the struck quark and the proton
spectators  which lead to diffractive events also  can produce single-spin asymmetries
(the Sivers effect) in semi-inclusive deep inelastic lepton scattering which survive in
the Bjorken limit.  In contrast to the SSAs arising from transversity and the Collins
fragmentation function, the fragmentation of the quark into hadrons is not necessary; one
predicts a correlation with the production plane of the quark jet itself $\vec S_p \cdot
\vec q \times \vec p_q.$

The final-state interaction mechanism provides an appealing physical
explanation within QCD
of single-spin asymmetries.  Remarkably, the same matrix element which
determines the
spin-orbit correlation
$\vec S \cdot \vec L$  also
produces the anomalous magnetic moment of the proton, the Pauli form
factor, and the
generalized parton distribution $E$ which is measured in deeply virtual Compton
scattering.  Physically, the final-state interaction phase arises as the
infrared-finite
difference of QCD Coulomb phases for hadron wave functions with differing
orbital angular
momentum.  An elegant discussion of the Sivers effect including its sign
has been
given by Burkardt~\cite{Burkardt:2004vm}.

The final-state interaction effects can also be identified with the gauge
link which
is present in the gauge-invariant definition of parton
distributions~\cite{Collins}.  Even when
the light-cone gauge is chosen, a transverse gauge link is required.  Thus
in any gauge
the parton amplitudes need to be augmented by an additional eikonal factor
incorporating
the final-state interaction and its phase~\cite{Ji:2002aa,Belitsky:2002sm}.
The net
effect is that it is possible to define transverse momentum dependent
parton distribution
functions which contain the effect of the QCD final-state interactions.

A related analysis also predicts that the initial-state
interactions from gluon exchange between the incoming quark and
the target spectator system lead to leading-twist single-spin
asymmetries in the Drell-Yan process $H_1 H_2^\updownarrow \to
\ell^+ \ell^- X$ \cite{Collins:2002kn,BHS2}.    Initial-state
interactions also lead to a $\cos 2 \phi$ planar correlation in
unpolarized Drell-Yan reactions \cite{Boer:2002ju}.

\subsection{Calculations of Single-Spin Asymmetries in QCD}

Hwang, Schmidt and I have calculated \cite{Brodsky:2002cx} the
single-spin Sivers asymmetry in semi-inclusive electroproduction
$\gamma^* p^{\updownarrow} \to H X$ induced by final-state
interactions in a model of a spin-\half~688 proton of mass $M$ with
charged spin-\half~and spin-0 constituents of mass $m$ and
$\lambda$, respectively, as in the QCD-motivated quark-diquark model
of a nucleon.  The basic electroproduction reaction is then
$\gamma^* p \to q (qq)_0$.  In fact, the asymmetry comes from the
interference of two amplitudes which have different proton spin, but
couple to the same final quark spin state, and therefore it involves
the interference of tree and one-loop diagrams with a final-state
interaction.  In this simple model the azimuthal target single-spin
asymmetry $A^{\sin \phi}_{UT}$ is given by
\begin{eqnarray}
A^{\sin \phi}_{UT} &=& {C_F \alpha_s(\mu^2) } \ { \Bigl(\ \Delta\,
M+m\ \Bigr)\ r_{\perp}\over \Big[\ \Bigl( \ \Delta\, M+m\
\Bigr)^2\
+\ {\vec r}_{\perp}^2\ \Big]}\nonumber \\
&\times& \Bigg[\ {\vec r}_{\perp}^2+\Delta
(1-\Delta)(-M^2+{m^2\over\Delta} +{\lambda^2\over 1-\Delta})\
\Bigg] \nonumber\\[1ex] &\times& \ {1\over {\vec r}_{\perp}^2}\
{\rm ln}{{\vec r}_{\perp}^2 +\Delta
(1-\Delta)(-M^2+{m^2\over\Delta}+{\lambda^2\over 1-\Delta})\over
\Delta (1-\Delta)(-M^2+{m^2\over\Delta}+{\lambda^2\over
1-\Delta})}\ . \label{sa2b}
\end{eqnarray}
Here $r_\perp$ is the magnitude of the transverse momentum of the
current quark jet relative to the virtual photon direction, and
$\Delta=x_{Bj}$ is the usual Bjorken variable.  To obtain
(\ref{sa2b}) from Eq. (21) of \cite{Brodsky:2002cx}, we used the
correspondence ${|e_1 e_2|/ 4 \pi} \to C_F \alpha_s(\mu^2)$
and the fact that the sign of the charges $e_1$ and $e_2$ of the
quark and diquark are opposite since they constitute a bound
state.  The result can be tested in jet production using an
observable such as thrust to define the  momentum $q + r$ of the
struck quark.

\vspace{0.3cm}
\begin{figure}[htp]
\begin{center}
\leavevmode {\epsfysize=10cm \epsffile{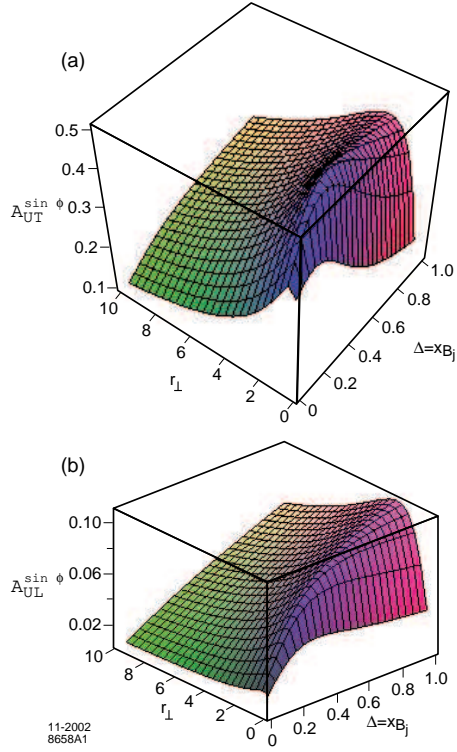}} 
\end{center}
\caption[*]{\baselineskip 13pt
 Model predictions for the target single-spin asymmetry
$A^{\sin \phi}_{UT}$ for charged and neutral current deep inelastic
scattering resulting
from gluon exchange in the final state.  Here $r_\perp$ is the magnitude of the
transverse momentum of the outgoing quark relative to the photon or vector boson
direction, and $\Delta = x_{bj}$ is the light-cone momentum fraction of the
struck quark.
The parameters of the model are given in the text.  In (a) the target
polarization is
transverse to the incident lepton direction.  The asymmetry in (b) $A^{\sin
\phi}_{UL} =
K A^{\sin \phi}_{UT}$ includes a kinematic factor $K = {Q\over
\nu}\sqrt{1-y}$ for the
case where the target nucleon is polarized along the incident lepton
direction.  For
illustration, we have taken $K= 0.26 \sqrt x,$ corresponding to the
kinematics of the
HERMES experiment~\cite{Airapetian:1999tv} with $E_{lab} = 27.6 ~{\rm GeV}$
and $y =
0.5.$} \label{fig:SSA1}
\end{figure}

The predictions of our model for the asymmetry $A^{\sin
\phi}_{UT}$ of the  ${\vec S}_p \cdot \vec q \times \vec p_q$
correlation based on  Eq. ({\ref{sa2b}}) are shown in Fig.
\ref{fig:SSA1}.  As representative parameters we take $\alpha_s =
0.3$, $M =  0.94$ GeV for the proton mass,  $m=0.3$ GeV for the
fermion constituent and $\lambda = 0.8$ GeV for the spin-0
spectator.  The single-spin asymmetry $A^{\sin \phi}_{UT}$ is
shown as a function of $\Delta$ and $r_\perp$ (GeV).  The
asymmetry measured at HERMES~\cite{Airapetian:1999tv}
$A_{UL}^{\sin \phi} = K A^{\sin \phi}_{UT}$ contains a kinematic
factor $K = {Q\over \nu}\sqrt{1-y} = {\sqrt{2Mx\over
E}}{\sqrt{1-y\over y}}$ because the proton is polarized along the
incident electron direction.  The resulting prediction for
$A_{UL}^{\sin \phi}$ is shown in Fig. \ref{fig:SSA1}(b).  Note
that $\vec r = \vec p_q - \vec q$ is the momentum of the current
quark jet relative to the photon momentum.  The asymmetry as a
function of the pion momentum $\vec p_\pi$ requires a convolution
with the quark fragmentation function.

Since the same matrix element controls the Pauli form factor, the
contribution of each quark current to the SSA is proportional to
the contribution $\kappa_{q/p}$ of that quark to the proton
target's anomalous magnetic moment $\kappa_p = \sum_q e_q
\kappa_{q/p}$~\cite{Brodsky:2002cx,Burkardt:2004vm}.
Avakian~\cite{Avakian:2002td} has shown that the data from HERMES
and Jefferson laboratory could be accounted for by the above
analysis.
The HERMES collaboration has recently measured the SSA in pion electroproduction
using transverse target polarization~\cite{Airapetian:2004tw}.
The Sivers and Collins effects can be separated using planar correlations;
both contributions
are observed to
contribute, with values not in disagreement with theory expectations.

It should be emphasized that the Sivers effect occurs even for jet
production; unlike
transversity, hadronization is not required.  There is no Sivers effect in
charged current
reactions since the $W$ only couples to left-handed
quarks~\cite{Brodsky:2002pr}.

The corresponding single spin asymmetry for the Drell-Yan processes, such
as $\pi
p^{\leftrightarrow}\ ({\rm or}\ p p^{\leftrightarrow}) \to \gamma^* X\to
\ell^+\ell^- X$,
is due to initial-state interactions.  The simplest way to get the result
is applying
crossing symmetry to the SIDIS processes.  The result that the SSA in the
Drell-Yan
process is the same as that obtained in SIDIS, with the appropriate
identification of
variables, but with the opposite sign~\cite{Collins,BHS2}.

We can also consider the SSA of $e^+e^-$ annihilation processes such as
$e^+e^-\to
\gamma^* \to \pi {\Lambda}^{\leftrightarrow} X$.  The $\Lambda$ reveals its
polarization
via its decay $\Lambda \to p \pi^-$.  The spin of the $\Lambda$ is normal
to the decay
plane.  Thus we can look for a SSA through the T-odd correlation
$\epsilon_{\mu \nu \rho
\sigma} S^\mu_\Lambda p^\nu_\Lambda q^\rho_{\gamma^*} p^\sigma_{\pi}$.
This is related
by crossing to SIDIS on a $\Lambda$ target.

Measurements from Jefferson Lab~\cite{Avakian:2003pk} also show
significant beam single spin asymmetries in deep inelastic
scattering.  Afanasev and Carlson~\cite{Afanasev:2003ze} have
recently shown that this asymmetry is due to the interference of
longitudinal and transverse photoabsorption amplitudes which have
different phases induced by the final-state interaction between
the struck quark and the target spectators just as in the
calculations of Ref. \cite{Brodsky:2002cx}.  Their results are
consistent with the experimentally observed magnitude of this
effect.  Thus similar FSI mechanisms involving quark orbital
angular momentum appear to be responsible for both target and beam
single-spin asymmetries.

\section{Heavy Quark Components of the Proton Structure Function}

In the simplest treatment of deep inelastic scattering, nonvalence quarks
are produced
via gluon splitting and DGLAP evolution.  However, in the full theory,
heavy quarks are
multiply connected to the valence quarks~\cite{Brodsky:1980pb}.  In fact,
the multiple
interactions of the sea quarks produce an asymmetry of the strange and
anti-strange
distributions in the nucleon due to their different interactions with the
other quark
constituents.  A QED analogy is the distribution of $\tau^+$ and $\tau^-$
in a higher Fock
state of muonium
$\mu^+ e^-.$  The $\tau^-$ is attracted to the $\mu^+$ thus asymmetrically
distorting its
momentum distribution.

The probability for Fock states of a light hadron such as the proton to have
an extra heavy quark pair decreases as $1/m^2_Q$ in non-Abelian gauge
theory~\cite{Franz:2000ee,Brodsky:1984nx}.  The relevant matrix element is
the cube of the
QCD field strength $G^3_{\mu nu}.$ This is in striking contrast to abelian
gauge theory where
the relevant operator is $F^4_{\mu \nu}$ and the probability of intrinsic
heavy leptons in QED
bound state is suppressed as $1/m^4_\ell.$  The intrinsic Fock state
probability is maximized
at minimal off shellness.  The maximum probability occurs at $x_i = {
m^i_\perp / \sum^n_{j =
1} m^j_\perp}$; i.e., when the constituents have equal rapidity.   Thus the
heaviest
constituents have the highest light-cone momentum fractions $x$.  Intrinsic
charm thus
predicts that the charm structure function has support at large $x_{bj}$ in excess of DGLAP
extrapolations~\cite{Brodsky:1980pb}; this is in agreement with the EMC
measurements~\cite{Harris:1995jx}.  As discussed in the next section, the
diffractive
dissociation of the intrinsic charm Fock state leads to leading charm hadron
production and fast charmonium production in agreement with
measurements~\cite{Anjos:2001jr}.   The production cross section for the
double charmed
$\Xi_{cc}^+$ baryon~\cite{Ocherashvili:2004hi} and the production of double
$J/\psi's$
appears to be consistent with the dissociation and coalescence of double IC Fock
states~\cite{Vogt:1995tf,BGK}.  Intrinsic charm can also explain the
$J/\psi \to \rho
\pi$ puzzle~\cite{Brodsky:1997fj},  and it affects the extraction of
suppressed CKM
matrix elements in $B$ decays~\cite{Brodsky:2001yt}.  Intrinsic charm can
also enhance
the production probability of Higgs bosons at hadron colliders from
processes such as $g
c \to H c.$ It is thus critical for new experiments (HERMES, HERA, COMPASS) to
definitively establish the phenomenology of the charm structure function at
large
$x_{bj}.$

\section{Diffraction Dissociation as a Tool to Resolve Hadron Substructure}

Diffractive multi-jet production in heavy nuclei provides a novel way to
measure the
shape of light-front Fock state wave functions and test color
transparency~\cite{Brodsky:1988xz}.  For example, consider the
reaction~\cite{Bertsch:1981py,Frankfurt:1999tq} $\pi A \rightarrow {\rm
Jet}_1 + {\rm
Jet}_2 + A^\prime$ at high energy where the nucleus $A^\prime$ is left
intact in its
ground state.  The transverse momenta of the jets balance so that $ \vec
k_{\perp i} +
\vec k_{\perp 2} = \vec q_\perp < {R^{-1}}_A \ . $ The light-cone
longitudinal momentum
fractions also need to add to $x_1+x_2 \sim 1.$   Diffractive dissociation
on a nucleus
also requires that the energy of the beam has to be sufficiently large such
that the
momentum transfer to the nucleus $\Delta p_L = {\Delta M^2\over 2 E_{lab}}$
is smaller
than the inverse nuclear size $R_A.$    The process can then occur
coherently in the
nucleus.

Because of color transparency, the valence wave function of the pion with
small impact
separation will penetrate the nucleus with minimal interactions,
diffracting into jet
pairs~\cite{Bertsch:1981py}.  The $x_1=x$, $x_2=1-x$ dependence of the di-jet
distributions will thus reflect the shape of the pion valence light-cone
wave function in
$x$; similarly, the $\vec k_{\perp 1}- \vec k_{\perp 2}$ relative
transverse momenta of
the jets gives key information on the second transverse momentum derivative
of the
underlying shape of the valence pion
wavefunction~\cite{Frankfurt:1999tq,Nikolaev:2000sh}.  The diffractive
nuclear amplitude
extrapolated to $t = 0$ should be linear in nuclear number $A$ if color
transparency is
correct.  The integrated diffractive rate will then scale as $A^2/R^2_A
\sim A^{4/3}.$ This is
in fact what has been observed by the E791 collaboration at FermiLab for
500 GeV incident
pions on nuclear targets~\cite{Aitala:2000hc}.  The measured momentum
fraction distribution
of the jets is found to be approximately consistent with the shape of the
pion asymptotic
distribution amplitude,
$\phi^{\rm asympt}_\pi (x) = \sqrt 3 f_\pi x(1-x)$~\cite{Aitala:2000hb}.
Data from
CLEO~\cite{Gronberg:1998fj} for the $\gamma
\gamma^* \rightarrow \pi^0$ transition form factor also favor a form for
the pion
distribution amplitude close to the asymptotic solution to its perturbative
QCD evolution
equation~\cite{Lepage:1979zb,Efremov:1978rn,Lepage:1980fj}.

The concept of high energy diffractive dissociation can be generalized to
provide a tool
to materialize the individual Fock states of a hadron or photon.  For
example, the
diffractive dissociation of a high energy proton on a nucleus $p A \to X
A^\prime$ where
the diffractive system is three jets $X= q q q$ can be used to determine
the valence
light-front wavefunction of the proton.

\subsection{Diffractive Dissociation and Hidden Color in Nuclear Wavefunctions}

In the case of a deuteron projectile, one can study diffractive processes such as $d A
\to p n A^\prime$ or $d A \to \pi^- p p$ to measure the mesonic Fock state
of a nuclear
wavefunction.  At small hadron transverse momentum, diffractive
dissociation of the deuteron should be
controlled by conventional nuclear interactions; however at large relative
$k_T$, the diffractive system should be sensitive to ``hidden color"
components of the
deuteron wavefunction.

In general, the six-quark wavefunction of a deuteron is a mixture of
five different color-singlet states.  The dominant color configuration at
large distances
corresponds to the usual proton-neutron bound state where transverse
momenta are  of
order ${\vec k}^2 \sim 2 M_d \epsilon_{BE}.$  However, at small impact
space separation,
all five Fock color-singlet components eventually acquire equal weight,
i.e., the
deuteron wavefunction evolves to 80\% hidden color.
At high $Q^2$ the deuteron form factor is sensitive to
wavefunction configurations where all six quarks overlap within an impact
separation
$b_{\perp i} < {\cal O} (1/Q).$  The derivation of the evolution
equation for the deuteron distribution amplitude and its leading anomalous
dimension
$\gamma$ is given in Ref.~\cite{Brodsky:1983vf}.  The relatively large
normalization of
the deuteron form factor observed at large $Q^2$~\cite{Farrar:1991qi}, as
well as the
presence of two mass scales in the scaling behavior of the reduced deuteron form
factor~\cite{Brodsky:1976rz} $f_d(Q^2)= F_d(Q^2)/F^2(Q^2/4)$ suggests sizable
hidden-color contributions in the deuteron wavefunction.

\vspace{0.3cm}
\begin{figure}[ht]
\begin{center}
\leavevmode {\epsfysize=4cm \epsffile{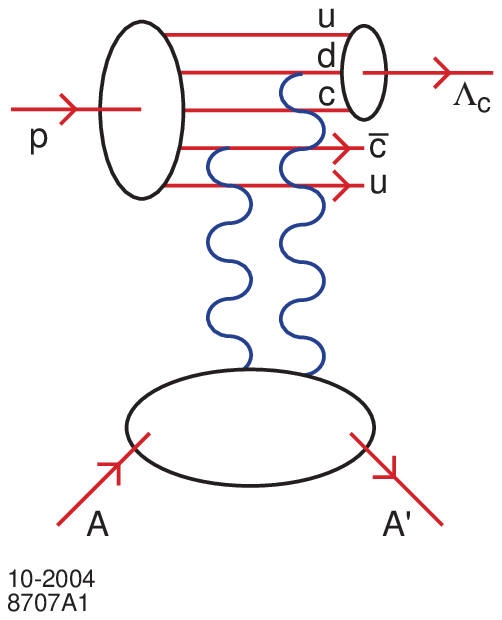}} 
\end{center}
\caption[*]{\baselineskip 13pt Production of forward heavy baryons by
diffractive
dissociation.} \label{DD}
\end{figure}

\subsection{Diffractive Dissociation and Heavy Quark Production}

Diffractive dissociation is particularly relevant to the production of
leading heavy
quark states.  The projectile proton can be decomposed as a sum over all of
its Fock
state components.  The diffractive dissociation of the intrinsic charm
$|uud c \bar c>$
Fock state of the proton on a nucleus can produce a leading heavy
quarkonium state at
high $x_F = x_c + x_{\bar c}~$ in $p A \to J/\psi X A^\prime$ since the $c$
and $\bar c$ can
readily coalesce into the charmonium state.  Since the constituents of a
given intrinsic
heavy-quark Fock state tend to have the same rapidity, coalescence of
multiple partons
from the projectile Fock state into charmed hadrons and mesons is also
favored.  For
example, as illustrated in  fig. \ref{DD}, one can produce leading
$\Lambda_c$ at high
$x_F$ and low $p_T$ from the coalescence of the $u d c$ constituents of the
projectile IC
Fock state.  A similar coalescence mechanism was used in atomic physics to
produce
relativistic antihydrogen in $\bar p A$ collisions~\cite{Munger:1993kq}.
This phenomena
is important not only for understanding heavy-hadron phenomenology, but also for
understanding the sources of neutrinos in astrophysics
experiments~\cite{Halzen:2004bn}.

The charmonium state will be produced at small transverse momentum and high
$x_F$  with a
characteristic $A^{2/3}$ nuclear dependence.  This forward contribution is
in addition to
the $A^1$ contribution derived from the usual PQCD fusion at small $x_F.$
Because of
these two components, the cross section violates perturbative QCD
factorization for hard
inclusive reactions~\cite{Hoyer:1990us}.  This is consistent with the observed
two-component cross section for charmonium production observed by the NA3
collaboration at CERN~\cite{Badier:1981ci}.

The production cross section for the double-charm $\Xi_{cc}^+$
baryon~\cite{Ocherashvili:2004hi} and the production of $J/\psi$ pairs
appears to be
consistent with the diffractive dissociation and coalescence of double IC Fock
states~\cite{BGK,Vogt:1995tf}.  It is unlikely that the appearance of two
heavy quarks at
high $x_F$ could be explained by the ``color drag model" used in PYTHIA
simulations~\cite{Andersson:1983ia} in which the heavy quarks are accelerated from low to
high $x$ by the fast valence quarks.  It is also conceivable that the
observations~\cite{Bari:1991ty} of
$\Lambda_b$ at high
$x_F$ at the ISR in high energy $p p$  collisions could be due to the
diffractive
dissociation and coalescence of the ``intrinsic bottom" $|uud b \bar b>$
Fock states of the
proton.

\vspace{.5in} \centerline{\bf Acknowledgements}

\noindent It is a pleasure to thank Professors Bill Gary and Ken Barish and
the other
organizers of ISMD 2004 for their hospitality in Sonoma.  I also thank my
collaborators,
particularly Rikard Enberg, Fred Goldhaber, Paul Hoyer, Dae Sung Hwang,
Gunnar Ingelman, Marek Karliner, and Ivan Schmidt.  This work was
supported by the Department of Energy, contract No. DE-AC02-76SF00515.

\end {document}